\documentclass[a4paper]{jpconf}
\usepackage{graphicx}
\usepackage[square,comma,sort&compress,numbers]{natbib}
\usepackage{amsmath}
\usepackage{floatrow}
\usepackage{soul}
\begin{document}
\title{Spectral modelling of a H.E.S.S.-detected pulsar wind nebula}

\author{Carlo van Rensburg$^1$, Paulus Kr\"{u}ger$^1$ and Christo Venter$^1$}

\address{$^1$Centre for Space Research, North-West University, Potchefstroom Campus, Private Bag X6001, Potchefstroom, 2520}

\ead{21106266@nwu.ac.za}

\begin{abstract}
In the last decade ground-based Imaging Atmospheric Cherenkov Telescopes have discovered roughly 30 pulsar wind nebulae at energies above 100 GeV. We present first results from a leptonic emission code that models the spectral energy density of a pulsar wind nebula by solving the Fokker-Planck transport equation and calculating inverse Compton and synchrotron emissivities. Although models such as these have been developed before, most of them model the geometry of a pulsar wind nebula as that of a single sphere. We have created a time-dependent, multi-zone model to investigate changes in the particle spectrum as the particles diffuse through the pulsar wind nebula, as well as predict the radiation spectrum at different positions in the nebula. We calibrate our new model against a more basic previous model and fit the observed spectrum of G0.9+0.1, incorporating data from the High Energy Stereoscopic System as well as radio and X-ray experiments. 
\end{abstract}

\section{Introduction}
In the last decade ground-based Imaging Atmospheric Cherenkov Telescopes (IACT$_{\rm s}$) have discovered almost 150 very-high-energy (VHE; $E >$ 100 GeV) $\gamma$-ray sources. Roughly 30 of these are confirmed pulsar wind nebulae (PWNe), while other source classes include supernova remnants, active galactic nuclei, or unidentified sources \footnote{tevcat.uchicago.edu}.  A subset of the unidentified sources may eventually turn out to be PWNe. H.E.S.S.\ is currently the world's largest IACT and is located near the Gamsberg mountain in Namibia.  H.E.S.S.\ consists of five telescopes which were built in two phases.  In December 2003 Phase I consisting of four 12-m telescopes was completed.  The fifth 28-m H.E.S.S.\ II telescope has been operational since July 2012, making H.E.S.S.\ the largest and most sensitive ground-based $\gamma$-ray telescope in the world.

A PWN is a bubble of shocked relativistic particles, produced when a pulsar's relativistic wind interacts with its environment \cite{Gaensler06}. The conversion of a fast-moving particle wind into electromagnetic radiation happens by two main processes.  The leptons in the particle wind interact with the magnetic field of the nebula and this causes synchrotron radiation (SR) up to several keV.  The second process is when the high-energy leptons interact with low-energy photons through inverse Compton scattering (IC), boosting the photon energies up to $E>100$ GeV.  Due to these two effects the radio, $X$-ray, and VHE $\gamma$-ray emissions are tightly linked, as all three emerge from the same lepton population. The following are typical characteristics of a PWN~\cite{Venter_Cherenkov05}:
\begin{enumerate}
 \item A filled-centre morphology which is brightest at the centre and dimming in all directions towards the edges;
 \item A flat radio photon spectrum with an index between 0 and -0.3;
 \item A well organised internal magnetic field structure;
 \item A high level of linear polarisation at high radio frequencies;
 \item Evidence of particle re-acceleration;
 \item Evidence of synchrotron cooling which means that the size of the PWN decreases with increasing energy.
\end{enumerate}

There are many unanswered questions in PWN physics. Surprisingly, it was noted \cite{Kargaltsev2010} that the measured $\gamma$-ray luminosity (1-10 TeV) of the PWNe does not correlate with the spin-down energy of their embedded pulsars. On the other hand, it was also found \cite{Mattana09} that the $X$-ray luminosity is correlated with the pulsar spin-down energy. Furthermore, it is currently unknown whether there is any correlation between the TeV surface brightness of the PWNe and the spin-down energy of their pulsars. Due to this reason it is necessary to create a spatially dependent model to calculate the spectral energy density (SED) from the PWN. The spatial dependence will yield the flux as a function of the radius. We calibrate our model against a more basic previous model and fit the spectrum of a known PWN. We describe our model in Section~\ref{sec:Model}, and its calibration and spectral fitting to G0.9+0.1 in Section~\ref{sec:Calibration}. We discuss our conclusions in Section ~\ref{sec:Conclusion}.

\section{PWN Model}\label{sec:Model}
\subsection{Transport equation and injection spectrum}
We model the transport of charged particles in a PWN by solving a Fokker-Planck-type equation similar to the Parker equation \cite{Parker1965}.  This equation includes diffusion, convection, energy losses, as well as a particle source.  If we neglect spatial convection, we are left with \cite{Kopp2013} 
\begin{equation}
\frac{\partial N_{\rm{e}}}{\partial t} = Q - \frac{\partial}{\partial E_{\rm e}}(\dot{E}_{\rm e}N_{\rm{e}}) + \nabla \cdot(\vec{\kappa} \cdot \nabla N_{\rm{e}}),
\label{kopp1}
\end{equation}
with $Q$ the particle injection spectrum, $N_{\rm{e}}$ the particle spectrum, $\dot{E}_{\rm e}$ the particle energy loss rate, and $\vec{\kappa}$ the diffusion tensor.  We used a broken power law for the particle injection spectrum following \cite{VdeJager2007},
\begin{equation}
Q(E_{\rm{e}},t) = \left\{\begin{matrix}
Q_0(t)\left(\frac{E_{\rm{e}}}{E_{\rm{b}}}\right)^{\alpha_1} E_{\rm e}<E_{\rm{b}}\\ 
Q_0(t)\left(\frac{E_{\rm{e}}}{E_{\rm{b}}}\right)^{\alpha_2} E_{\rm e} \geq E_{\rm{b}}.
\end{matrix}\right.
\label{brokenpowerlaw}
\end{equation}  
Here $Q_0$ is the normalisation constant, $E_{\rm{b}}$ the break energy, $E_{\rm{e}}$ the lepton energy and $\alpha_1$ and $\alpha_2$ the spectral indices. To obtain $Q_0$ we use a spin-down luminosity $L(t) = L_0/\left(1+t/\tau_0\right)^2$ of the pulsar \cite{Reynolds1984}, with  $\tau_0$ the initial spin-down time scale of the pulsar, and $L_0$ the initial spin-down luminosity. We normalise $Q$ as follows:
\begin{equation}
  \epsilon L =  \int_{E_{\rm{min}}}^{E_{\rm{max}}}QE_{\rm e}dE_{\rm e},
  \label{normQ}
\end{equation}
with $\epsilon$ the conversion efficiency of the spin-down luminosity to particle power.
\subsection{Radiative energy losses in the PWN}
Particle energy is dissipated from the system due to radiation.  We incorporated SR and IC scattering, similar to calculations done by \citep{Kopp2013} in their globular cluster model. SR losses are given by \cite{BlGould1970}
\begin{equation}
  \left(\frac{dE_{\rm e}}{dt}\right)_{\rm{SR}} = -\frac{\sigma_{\rm T}c}{6 \pi}\frac{E_{\rm e}^2 B^2}{(m_{\rm e}c^2)^2},
  \label{SR}
\end{equation}
with $\sigma_T$ the Thompson cross section and $B$ the PWN magnetic field. The IC scattering energy loss rate is given by
\begin{equation}
  \left(\frac{dE_{\rm e}}{dt}\right)_{\rm{IC}} = -\frac{g_{\rm{IC}}}{E_{\rm{e}}^2} \sum_{p=1}^{3} \int \int n_{\varepsilon,p}(r,\varepsilon,T_p) \frac{E_\gamma}{\varepsilon} \hat{\zeta}(E_{\rm{e}},E_{\gamma},\varepsilon) d \varepsilon dE_\gamma,
  \label{IC}
\end{equation}
with $g_{\rm{IC}} = 2\pi e^4c$, $\varepsilon$ the soft photon energy, $n_{\varepsilon,p}$ the blackbody photon number density, $T_{p}$ the photon temperature of component $p$, $E_{\gamma}$ the TeV up-scattered photon energy, and $\zeta(E_{\rm{e}},E_{\gamma},\varepsilon) = \zeta_0 \hat{\zeta}(E_{\rm{e}},E_{\gamma},\varepsilon)$ the collision rate with $\zeta_0 = 2\pi e^4E_0c/\varepsilon E_{\rm{e}}^2$, and $\hat{\zeta}$ given by
\begin{equation}
\hat{\zeta}(E_{\rm{e}},E_{\gamma},\varepsilon) =\left\{\begin{matrix}
0 &\rm{if} &E_{\gamma} \leq \frac{\varepsilon E_0^2}{4 E_{\rm{e}}^2},\\ 
\frac{E_{\gamma}}{\varepsilon}-\frac{E_0^2}{4 E_{\rm{e}}^2} &\rm{if} &\frac{\varepsilon E_0^2}{4 E_{\rm{e}}^2} \leq E_{\gamma} \leq \varepsilon,\\ 
f(q,g_0) &\rm{if} &\varepsilon \leq E_{\gamma} \leq \frac{4\varepsilon E_{\rm{e}}^2}{E_0^2 + 4\varepsilon E_{\rm{e}}},\\ 
0 &\rm{if} &E_{\gamma} \geq \frac{4\varepsilon E_{\rm{e}}^2}{E_0^2 + 4\varepsilon E_{\rm{e}}}.
\end{matrix}\right.
\end{equation}
Here, $E_0 = m_{\rm e}c^2$, $f(q,g_0) = 2q$ln$q+(1-q)(1+(2+g_0)q)$, $q=E_0^2E_{\gamma}/(4\varepsilon E_{\rm{e}}(E_{\rm{e}}-E_{\gamma}))$, and $g_0(\varepsilon,E_{\gamma}) = 2\varepsilon E_{\gamma}/E_0^2$.
The two radiation loss rates can be added to find $\dot{E}_{\rm e}$ used in (\ref{kopp1}).

\subsection{Calculation of the particle spectrum}
If we use spherical coordinates and assume spherical symmetry, i.e. $\frac{\partial}{\partial \theta} = 0$ and $\frac{\partial}{\partial \phi} = 0$, the only change in particle (lepton) spectrum will be in the radial direction, so $\nabla^2 = \frac{1}{r^2}\left(\frac{\partial}{\partial r}\left(r^2\frac{\partial N}{\partial r}\right)\right).$ For the diffusion scalar coefficient $\kappa$ we consider Bohm diffusion
\begin{equation}
  \kappa = \kappa_B\frac{E_{\rm e}}{B},
\end{equation}
with $\kappa_B = c/3e$, and $c$ and $e$ denote the speed of light and the elementary charge. We can now discretise (\ref{kopp1}): 
\begin{equation}
\begin{split}
 (1-z+\beta)N_{i,j+1,k} = (1+z-\beta)N_{i,j-1,k} + (\beta+\gamma)N_{i,j,k+1} + (\beta-\gamma)N_{i,j,k-1} + \\ \frac{2}{\delta E_2 + \delta E_1}\left(r_a\dot{E}_{i+1,j,k}N_{i+1,j,k} - \frac{1}{r_a}\dot{E}_{i-1,j,k}N_{i-1,j,k} \right) + Q_{i,j,k}dt,
 \end{split}
 \label{fin_EQ}
\end{equation}
with $i$ the time index, $j$ the energy index, $k$ the radial index,  $\beta = 2\kappa dt/ \triangle r^2$, $\gamma = 2\kappa dt/(r\triangle r)$ with $\triangle r$ the size in the spatial bins. Also $r_{\rm a} = \triangle E_{i+1}/\triangle E_{i}$ with
\begin{equation}
z = \left(\frac{1}{\triangle E_{{i+1}}-\triangle E_{{i}}}\right)\left(\frac{1}{r_{{\rm a}}}-r_{{\rm a}}\right)\dot{E}_{{i,j,k}}.
\end{equation}

We initially approached the discretisation process by using a simple Euler method. It soon became clear that this method was not stable. We then decided to use a DuFort-Frankel scheme to solve (\ref{kopp1}). In solving this equation, we calculate the electron spectrum of the PWN due to the injected particles from the embedded pulsar, taking into account their diffusion through the PWN and the IC and SR energy losses. We use the following parametrised form for $B(t)$ \cite{VdeJager2007}
\begin{equation}
B(t) = \frac{B_0}{1+\left(\frac{t}{\tau_0}\right)^{\alpha_B}},
\label{B_Field}
\end{equation}
with $B_0$ the birth magnetic field, $t$ the age of the PWN, and $\alpha_B$ the magnetic field parameter. We can calculate $L_0$ and $B_0$ by using the present-day luminosity and the present-day magnetic field together with the age of the PWN. We limit the particle energy using $E_{\rm{max}} = \frac{e}{2}\sqrt{\frac{L(t)\sigma}{c(1+\sigma)}}$~\cite{VdeJager2007}, with $\sigma$ the ratio of electromagnetic to particle luminosity. Particles with $E_{\rm e}>E_{\rm{max}}$ are assumed to escape.

To solve (\ref{fin_EQ}) numerically, our multi-zone model divides the PWN into shells. The particles are injected into zone one close to the centre and are allowed to diffuse through the different zones. We assumed as initial condition, that all zones were devoid of any particles.  We used a reflective boundary condition at the inner boundary of the PWN, and on the outer rim we set $N_{\rm{e}}$ equal to zero to allow the electrons to escape from the PWN.  

\begin{table}[b]
\small{
\begin{tabular}{|lll|}
\hline
Model Parameter & Symbol &  Value\\  \hline \hline
Braking index & n & 3  \\
$B$-field parameter & $\alpha_B$ & 0.5 \\
Present day $B$-field & $B(T)$ & 40.0 $\mu G$ \\
Conversion efficiency  &    $\epsilon$    &  0.6\\
Age & T & 1900 $yr$\\
Initial period of the pulsar & $P_0$ & 43 ms\\
Current period of the pulsar & $P$ & 52 ms\\
Pulsar spin-down rate & $\dot{P}$ & $1.5557 \times 10^{-13}$ \\
Distance & d & 8.5 kpc  \\
Q index 1& $\alpha_1$& -1.0 \\
Q index 2& $\alpha_2$& -2.6 \\
Maximum energy parameter & $\sigma$ & 0.2 \\
Soft photon component 1 & $T_1$ and $u_1$ & $2.76\:{\rm K}$, $0.23\:{\rm eV/cm}^3$ \\
Soft photon component 2 & $T_2$ and $u_2$ & $35\:{\rm K}$, $0.5\: {\rm eV/cm}^3$ \\  
Soft photon component 3 & $T_3$ and $u_3$ & $4500\:{\rm K}$, $50\: {\rm eV/cm}^3$ \\ \hline
\end{tabular}
}
\caption{Values of model parameters.}
\label{tbl:param}
\end{table}

\subsection{Calculation of radiation spectrum}
The time-dependent photon spectrum of each zone can now be calculated, given the electron spectrum solved for each zone. For IC we have \citep{Kopp2013}
\begin{equation}
\left(\frac{dN_\gamma}{dE_\gamma}\right)_{IC} = \frac{g_{IC}}{A} \sum_{p=1}^{3} \int \int n_{\varepsilon,p}(r,\varepsilon,T_p) \frac{\mathcal{N}_{\rm{e}}}{\varepsilon E_{\rm{e}}^2}\hat{\zeta}(E_{\rm{e}},E_\gamma,\varepsilon) d\varepsilon dE_{\rm{e}} , 
\label{ICrad}
\end{equation}
where $A=4\pi d^2$, $d$ the distance to the source, and $\mathcal{N}_{\rm{e}}$ is the number of electrons per energy in a spherical shell around $r$. We consider three components of target photons, cosmic background radiation (CMB) with $T_1 = $ 2.76 K and $u_1 = $ 0.23 eV/cm$^3$, Galactic background infra-red photons with $T_2 = $ 35 K and $u_2 =$ 0.5 eV/cm$^3$, and starlight with $T_3 =$ 4 500 K and $u_3 =$ 50 eV/cm$^3$. For SR we have
\begin{equation}
\left(\frac{dN_\gamma}{dE_\gamma}\right)_{\rm{SR}} = \frac{1}{A}\frac{1}{hE_\gamma}\frac{\sqrt{3}e^3B(r)}{E_0} \int \int_0^{\pi/2} \mathcal{N}_{\rm{e}}(E_{\rm{e}},r)\tilde{\kappa} \left(\frac{\nu}{\nu_{\rm{cr}}(E_{\rm{e}},\theta,r)}\right)\sin^2 \theta d\theta dE_{\rm{e}},
\label{SRrad}
\end{equation}
with $\nu_{\rm{cr}}$ the critical frequency (with pitch angle $\theta$) 
\begin{equation}
\nu_{\rm{cr}}(E_{\rm{e}},\theta,r) = \frac{3ec}{4\pi E_0^3}E_{\rm{e}}^2 B(r)\sin\theta
\end{equation}
and $\tilde{\kappa}$ given by
\begin{equation}
\tilde{\kappa}(x) = x\int_x^\infty K_{5/3}(y)dy,
\end{equation}
where $K_{5/3}$ a modified Bessel function of the second kind of order $5/3$.

\subsection{Calculation of the line of sight flux}
In order to do the line of sight (LOS) integration of the flux from the PWN we need to use the flux per unit volume calculated in the previous section and multiply it with the volume in a particular LOS as viewed from Earth.  Since our model is spherically symmetric and since the source is very far from Earth, we used cylinders intersecting the spherical zones. We chose the cylinders and the spheres to have the same radii. We then used the intersection volume to calculate the flux in a particular LOS. We will present the radial dependence of the flux elsewhere.

\begin{figure}[t]
\begin{minipage}[b]{5in}
\includegraphics[angle = -90,width=5in]{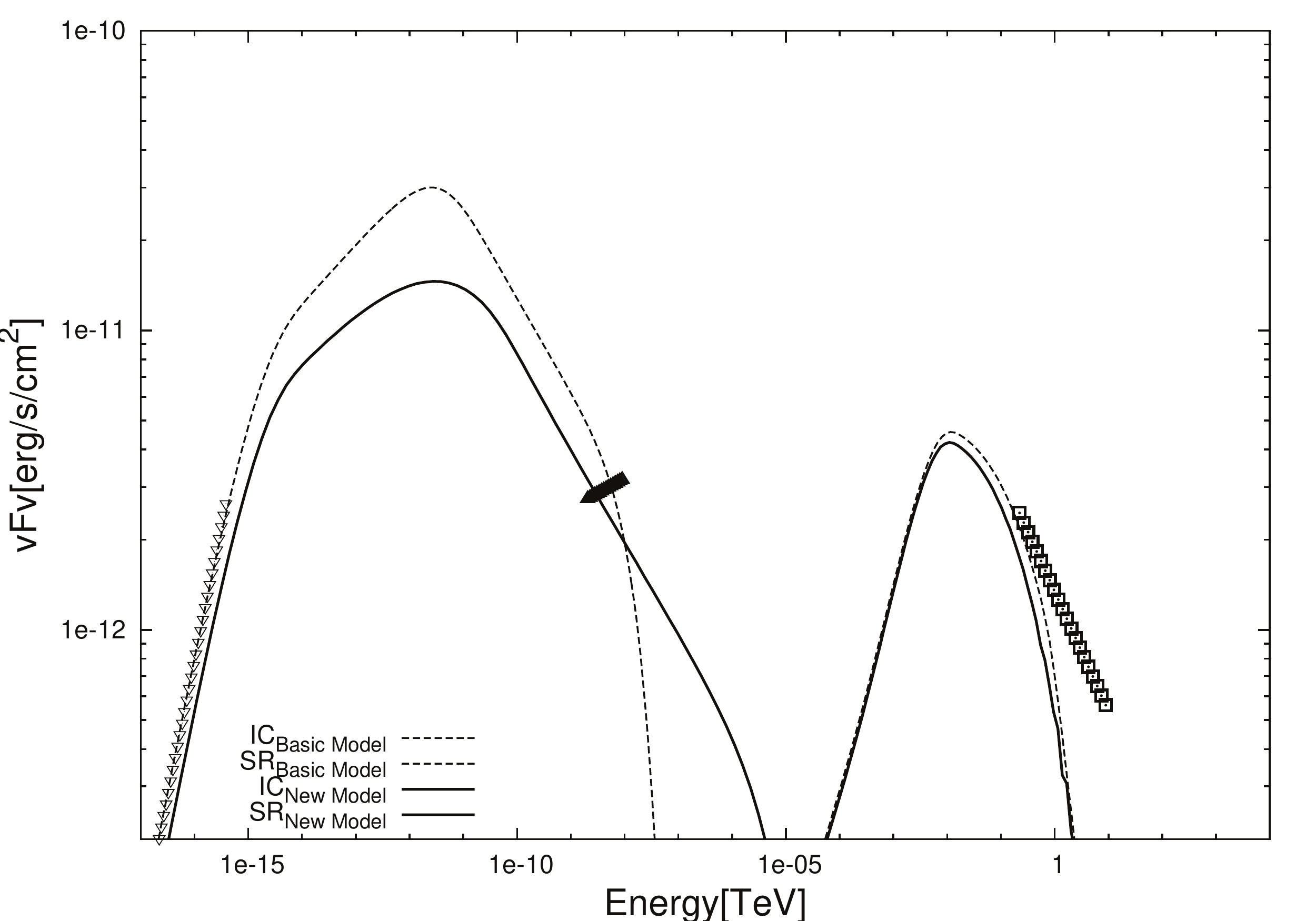}
\caption{\label{fig:Calibrate}Calibration of our model against the model of \cite{VdeJager2007}. The radio data are from \cite{HelfandB1987}, the X-ray data are from \cite{Sidoli2004}, and the $\gamma$-ray data are from \cite{Porquet2003}.}
\end{minipage} 
\end{figure} 

\section{Model calibration and spectral modelling of G0.9+0.1}\label{sec:Calibration}
To calibrate our model we compared our results to those of \cite{VdeJager2007} as shown in figure~\ref{fig:Calibrate}. Pulsar J1747$-$2809 has now been discovered in PWN G0.9+0.1 \cite{G0.9+0.1data} with $P = 52$ ms and $\dot{P} = 1.5557\times10^{-13}$. We used these values to calculate $\tau_0 = P_0/2\dot{P_0}$ assuming $P_0 = 43$ ms and no decay of the pulsar $B$-field. In the figure, the radio data are from \cite{HelfandB1987}, the $X$-ray data are from \cite{Sidoli2004}, and the $\gamma$-ray data are from \cite{Porquet2003}. We list our model parameters in table \ref{tbl:param}.

\section{Conclusions}\label{sec:Conclusion}
We created a time-dependent, multi-zone SED model to model the radiation spectrum observed from PWNe. We calibrated this model using a previous model. We then fitted the data of G0.9+0.1. The new model does not exactly reproduce \cite{VdeJager2007}, but the results are quite close. The new SR component is lower due to the fact that the old model did not consider IC losses while performing the particle-transport. We will next perform a population study to probe a potential relationship between the TeV surface brightness and the spin-down luminosity of the embedded pulsar. We also intend to investigate the spatial dependence of the different fluxes.

\bibliography{Bibliography}

\providecommand{\newblock}{}
\begin{thebibliography}{10}
\expandafter\ifx\csname url\endcsname\relax
  \def\url#1{{\tt #1}}\fi
\expandafter\ifx\csname urlprefix\endcsname\relax\def\urlprefix{URL }\fi
\providecommand{\eprint}[2][]{\url{#2}}

\bibitem{Gaensler06}
{Gaensler} B~M and {Slane} P~O 2006 {\em ARA\&A\/} {\bf 44} 17--47

\bibitem{VdeJager2007}
{Venter} C and {de Jager} O~C 2007 {\em WE-Heraeus Seminar on Neutron Stars and
  Pulsars 40 years after the Discovery\/} ed {Becker} W and {Huang} H~H p~40

\bibitem{Kargaltsev2010}
{Kargaltsev} O and {Pavlov} G~G 2010 {\em X-ray Astronomy 2009; Present Status,
  Multi-Wavelength Approach and Future Perspectives\/} {\bf 1248} 25--28

\bibitem{Mattana09}
{Mattana} F, {Falanga} M, {G{\"o}tz} D, {Terrier} R, {Esposito} P, {Pellizzoni}
  A, {De Luca} A, {Marandon} V, {Goldwurm} A and {Caraveo} P~A 2009 {\em ApJ\/}
  {\bf 694} 12--17

\bibitem{Parker1965}
{Parker} E~N 1965 {\em ApJ\/} {\bf 142} 1086

\bibitem{Kopp2013}
{Kopp} A, {Venter} C, {B{\"u}sching} I and {de Jager} O~C 2013 {\em ApJ\/} {\bf
  779} 126

\bibitem{Venter_Cherenkov05}
{de Jager} O and {Venter} C 2005 {\em Towards a Network of Atmospheric
  Cherenkov Detectors VII (astro-ph/0511098)\/} ed {Degrange} B and {Fontaine}
  G (\textit{Preprint} \eprint{arXiv:astro-ph/0511098})

\bibitem{Reynolds1984}
{Reynolds} S~P and {Chevalier} R~A 1984 {\em apj\/} {\bf 278} 630--648

\bibitem{BlGould1970}
{Blumenthal} G~R and {Gould} R~J 1970 {\em Reviews of Modern Physics\/} {\bf
  42} 237--271

\bibitem{G0.9+0.1data}
{Camilo} F, {Ransom} S~M, {Gaensler} B~M and {Lorimer} D~R 2009 {\em ApJ\/}
  {\bf 700} L34--L38

\bibitem{HelfandB1987}
{Becker} R~H and {Helfand} D~J 1987 {\em ApJ\/} {\bf 316} 660--662

\bibitem{Sidoli2004}
{Sidoli} L, {Bocchino} F, {Mereghetti} S and {Bandiera} R 2004 {\em MmSAI\/}
  {\bf 75} 507

\bibitem{Porquet2003}
{Porquet} D, {Decourchelle} A and {Warwick} R~S 2003 {\em A\&AS\/} {\bf 401}
  197--203

\end{thebibliography}
\end{document}